\documentstyle[preprint,prd,aps,epsf]{revtex}

\input{psfig}
\input{epsf.tex}

\newcommand {\Dtau}{\Delta\tau}

\begin{document}
\draft

\title{\bf
Mass gap in compact U(1) Model in (2+1) dimensions}

\author{Mushtaq Loan\footnote{e-mail :mushe@newt.phys.unsw.edu.au},
Michael Brunner  and Chris Hamer}

\address{School of Physics, University of New South Wales, Sydney, Australia}

\date{September 25, 2002}
%commat out date
\maketitle
\begin{abstract}
A numerical study of low-lying glueball masses of compact U(1) lattice 
gauge theory in (2+1) dimensions is performed 
using Standard Path integral Monte Carlo techniques. The masses are
extracted, at fixed (low) temperature, from simulations on 
anisotropic lattices, with temporal
lattice spacing much smaller than the spatial ones. Convincing evidence of
the scaling behaviour in the antisymmetric mass gap is observed 
over the range $1.4<\beta <2.25$. The observed behaviour is very consistent 
with asymptotic form predicted by 
G{\" o}pfert and Mack. 
Extrapolations are made to the  ``Hamiltonian" limit, and 
the results are compared with  previous estimates obtained 
by many other Hamiltonian studies.
\end{abstract}
\vspace{10mm}

\section{Introduction}
Compact U(1) lattice gauge theory in three-dimensions is the most 
well-understood nontrivial gauge theory. The theory exhibits some
important similarities with QCD \cite{ham94}, such as confinement 
\cite{pol78,pol77} and chiral symmetry breaking \cite{fie90}.
Because of its simplicity, it has been a 
good testing
ground for  new methods and algorithmic approaches.                                               
While the Euclidean Path Integral Monte Carlo simulations \cite{cre79} 
have been 
very successful in the study of non-perturbative lattice gauge theories 
and is currently the preffered technique for studying QCD at low 
energies, there  there has been 
some progress in the 
development of analytic and numerical approaches in the Hamiltonian 
formalism. 
The Hamiltonian version of the model has been studied by many methods:
some recent studies include series expansions
\cite{ham92}, finite-lattice techniques\cite{irv83}, the t-expansion
\cite{hor87,mor92}, and coupled-cluster techniques
\cite{dab91,fan96,bak96},
 as well
as Quantum Monte Carlo 
methods\cite{chi84,koo86,yun86,ham94,john97,cher01,ham00}.
Quite accurate estimates have been obtained for the string
tension and the mass gaps, which can be used as comparison for our
present results. The finite-size scaling    
properties of the model can be predicted using an effective Lagrangian
approach combined with a weak-coupling expansion
\cite{ham93}, and the predictions agree very well with finite-lattice data
\cite{ham94}.     
Here our aim is to use standard Euclidean path integral Monte Carlo 
techniques for an anisotropic lattice, and see whether useful results can 
be obtained in the anisotropic or Hamiltonian limit.

\section{Compact U(1) Model in 3 Dimensions}
The anisotropic Wilson gauge action for compact U(1) model in (2+1) 
dimensions has the form \cite{mor97}
\begin{equation}
S = \beta_{s}\sum_{r,i<j}P_{ij}(r)+\beta_{t}\sum_{r,i}P_{it}(r)
\label{eqn1}
\end{equation}
Where $P_{ij}$ and $P_{it}$ are the spatial and temporal plaquette 
variables respectively. In the classical limit 
\begin{eqnarray}
\beta_{s} & = & \frac{a_{t}}{e^{2}a_{s}^{2}}  = \frac{1}{g^{2}}\Dtau
\nonumber\\
 \beta_{t} & = & \frac{1}{e^{2}a_{t}}  = \frac{1}{g^{2}}\frac{1}{\Dtau}
\label{eqn3}
\end{eqnarray}
 where $\beta =1/g^{2}$ 
($g^{2}=ae^{2}$ in (2+1)D)) and $\Dtau = 
a_{t}/a_{s} $ is the anisotropy
parameter, $a_{s}$ is the lattice
 spacing in the space direction, and $a_{t}$
is the temporal spacing.
In the weak-coupling approximation, the above action can be written as
 \begin{equation}
 S  = \beta \bigg[\Dtau \sum_{r}\sum_{i<j}\bigg(1-\cos 
\theta_{ij}(r)\bigg)
+\frac{1}{\Dtau }\sum_{r,i}\bigg(1-\cos \theta_{it}(r)\bigg)\bigg]
\label{eqn4}
\end{equation} 
In the limit $\Dtau \rightarrow 0$, the time variable becomes continuous 
and one obtains the Hamiltonian limit of the model (modulo a Wick rotation 
back to Minkowski space).

Among other features, antisymmetric mass gap is 
a fundamental quantity of interest in  U(1) model. On an isotropic 
lattice, the model  reduces to the simple continuum theory of 
non-interacting photons in the naive continuum limit at a fixed energy 
scale \cite{gro83} but if one renormalizes or rescales in the standard
way so as to maintain the
mass gap constant, then one obtains a confining theory of free massive 
bosons. G{\" o}pfert and Mack\cite{gop82} proved that in the continuum 
limit the theory converges to a scalar free field theory of massive 
bosons.   
They found that in that limit the mass gap behaves as
\begin{equation}
am_D = \sqrt{\frac{8\pi^2}{g^2}}\exp(-\frac{\pi^2}{g^2}v(0))
\label{eqn5}
\end{equation} 
where $v(0)=0.2527$ is the Coulomb potential at zero distance for the 
isotropic case.

The behaviour of the mass gap in the anisotropic case will be similar
to equation (\ref{eqn5}) Generalizing
discussions by Banks {\it et al}\cite{ban77} and
Ben-Menahem\cite{ben79}, we find that the exponential
factor takes exactly the same form in the anisotropic case.
The only difference is that the lattice Coulomb potential at zero
spacing for general $\Delta \tau$ is   
\begin{eqnarray}
v(0) & = & \int^{\pi}_{-\pi} \frac{d^{3}k}{(2\pi)^{3}} \frac{\Delta
\tau}{4[\sin^{2}(k_0/2) + \Delta \tau^{2}(\sin^{2}(k_1/2) +
\sin^{2}(k_2/2))]} \\
 & = & \left\{ \begin{array} {c}
 0.2527 \hspace{5mm} (\Delta \tau = 1) \\
 0.3214 \hspace{5mm} (\Delta \tau = 0)
\end{array}
\right.
\label{eqn7}
\end{eqnarray}
The above result is based on dilute gas approximation,
which is justified in the
Euclidean case, but not in the Hamiltonian limit\cite{ban77,ben79}.

The expected finite-size scaling behaviour of the mass gap near the
continuum critical point in this model
is not known; but Weigel and Janke\cite{wei99} have performed a Monte
Carlo simulation for an O(2) spin model in three dimensions which should
lie in the same universality class, obtaining

\begin{equation}
M \sim 1.3218/L
\label{eqn8}
\end{equation}      
for the magnetic gap.

\section{Monte Calro simulations and glueball masses}
Using the anisotropic Wilson action, we perform 
Monte Carlo simulations on a finite lattice of size $N_{s}^{2}\times
N_{\tau}$, where
$N_{s}$ is the number of lattice sites in the space direction and
$N_{\tau}$
in the temporal direction, with spacing ratio
$\Delta \tau = a_{t}/a_{s} $.
By varying $\Delta \tau $ it is possible to change $a_{t}$, while
keeping the spacing in the spatial direction fixed.
The details of the algorithm for updating are given elsewhere \cite{loana}.
The
simulations were performed on lattices with $N_s=16$ sites in each of
the two
spatial directions and
$N_{t}= 16 - 64$ in the temporal direction for a range
of couplings $\beta = 1 - 3$.

\section{Results and Discussion}

\begin{figure}
\centerline{\psfig{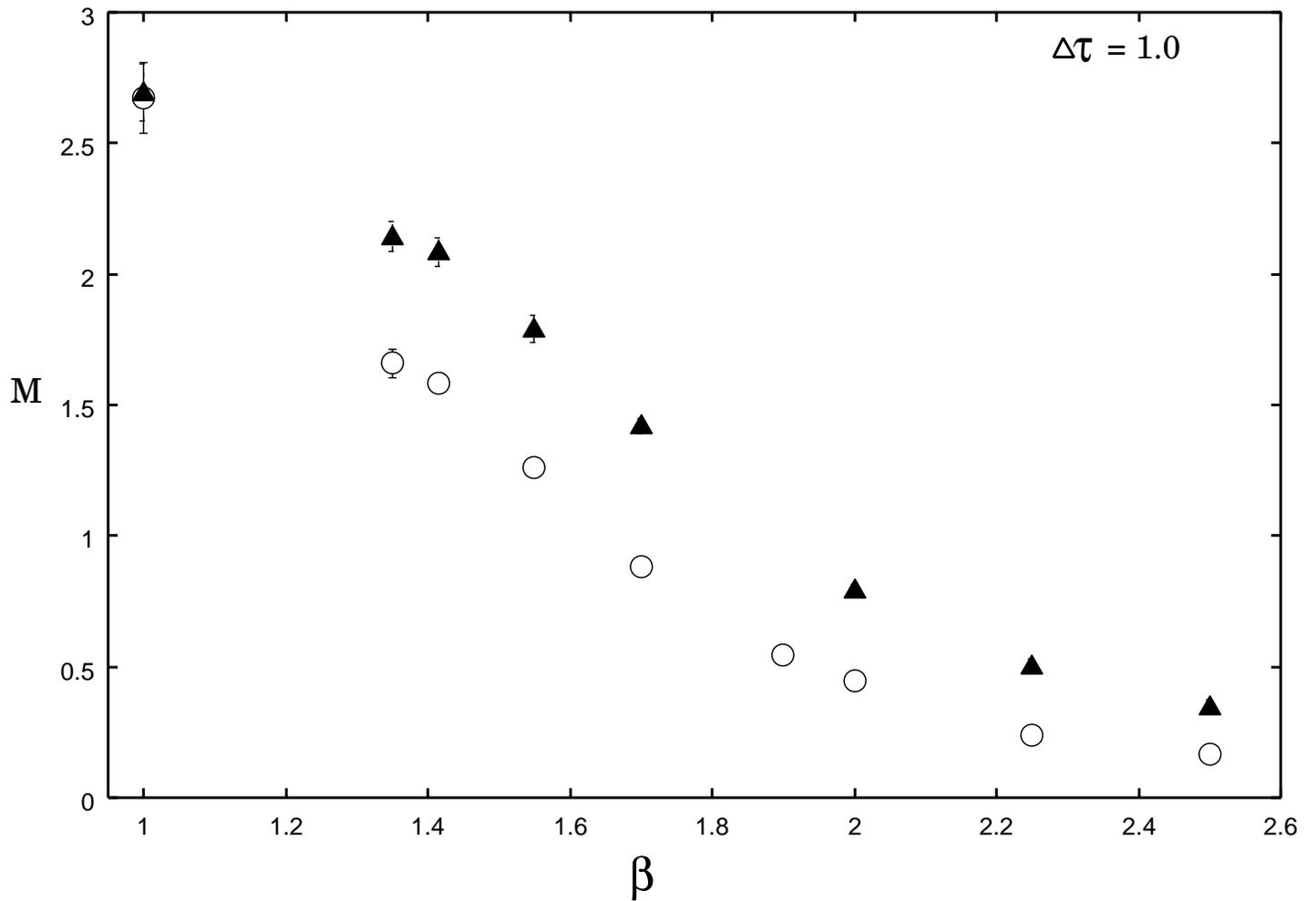}}
\caption{The glueball mass estimates of 
as a function of $\beta$. The estimates for $0^{++}$ and $0^{--}$ channels
are shown by solid triangles and open circles respectively.}
\label{fig1}
\end{figure}

\begin{figure}
\centerline{\psfig{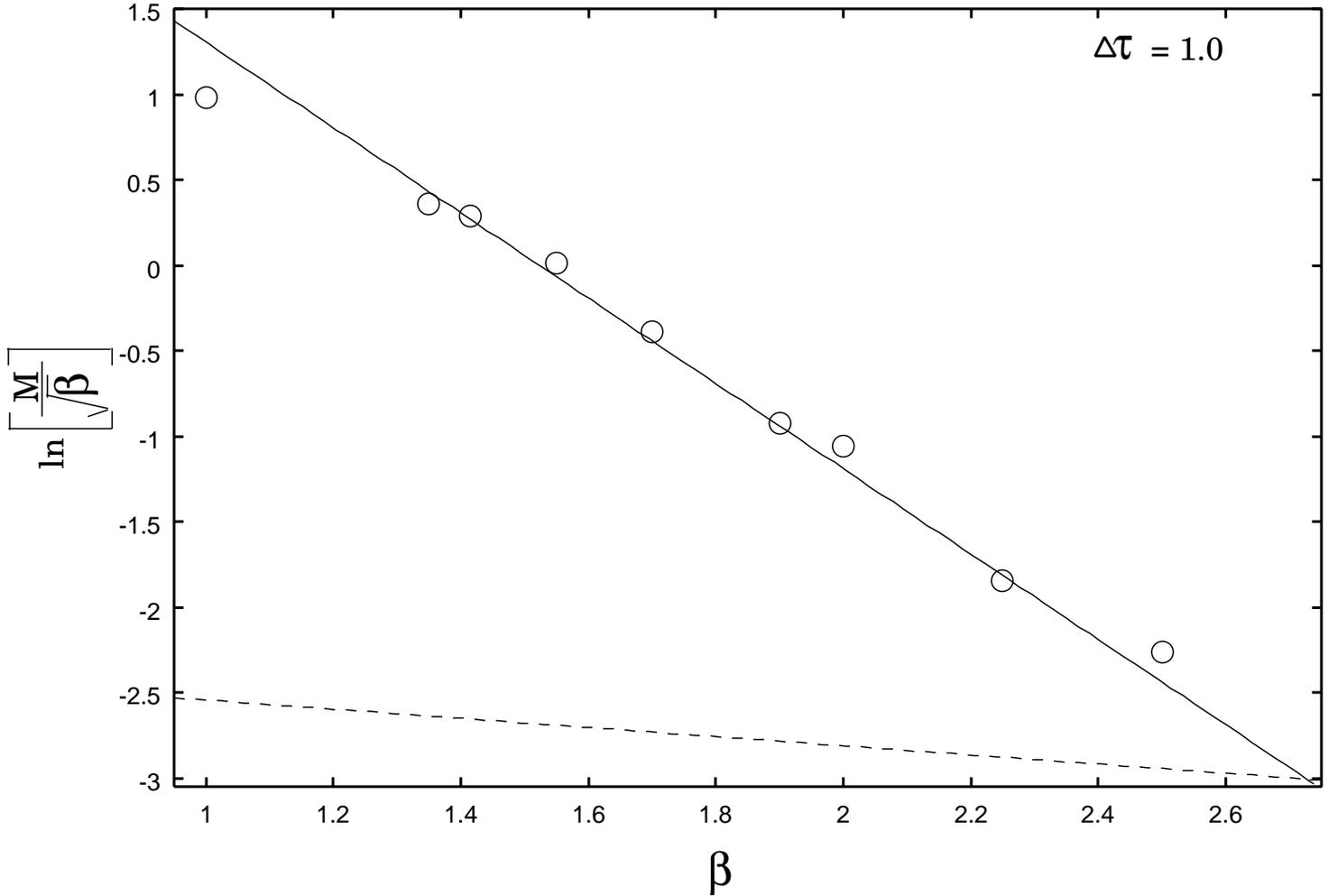}}
\caption{The scaling behaviour of the antisymmetric mass gap against 
$\beta$
at $\Delta \tau =
1.0$. The solid line is a fit of the form eq. (\protect\ref{eqn5}). The 
errors
are smaller than the symbols.
The dashed line  shows the finite size scaling behaviour 
\protect\cite{wei99}}.
\label{fig2}
\end{figure}

Figure \ref{fig1} shows the  estimates for the  glueball masses of
$0^{++}$
and $0^{--}$ channels
against $\beta$ for the Euclidean case  $\Dtau =1.0$. As
expected the  masses are
seen to
vanish exponentially with $\beta$ as we approach the continuum limit
(high $\beta$).
To show the evidence of  scaling behaviour of antisymmetric glueball
masses , we
used the predicted  asymptotic form, equation \ref{eqn5}, with
an additional normalization constant;
\begin{equation}
M=am_{D}=c_{1}\sqrt{8\pi^{2}\beta}\mbox{exp}\bigg(-\pi^{2}\beta v(0)\bigg)
\label{eqn9}
\end{equation}
where $c_{1}=5.23\pm 0.11$ when adjusted to  fit the data. Figure   
\ref{fig2} shows the predicted scaling behaviour. The solid line on the
graph is a fit to the data over the range $1.4\leq \beta \leq 2.25$. The
slope, $2.48\pm 0.09$, of the data matches the predicted asymptotic form
very nicely, but the coefficient is too large by a factor of 5.2. It would
again be interesting to explore whether this discrepancy could be due to
quantum corrections. The finite-size scaling behaviour is shown by the                                                  
dashed line in Figure \ref{fig2}.
It can be seen that the Euclidean mass gap should
not be effected by finite-size corrections until $\beta \geq 2.8$.

\begin{figure}
\centerline{\psfig{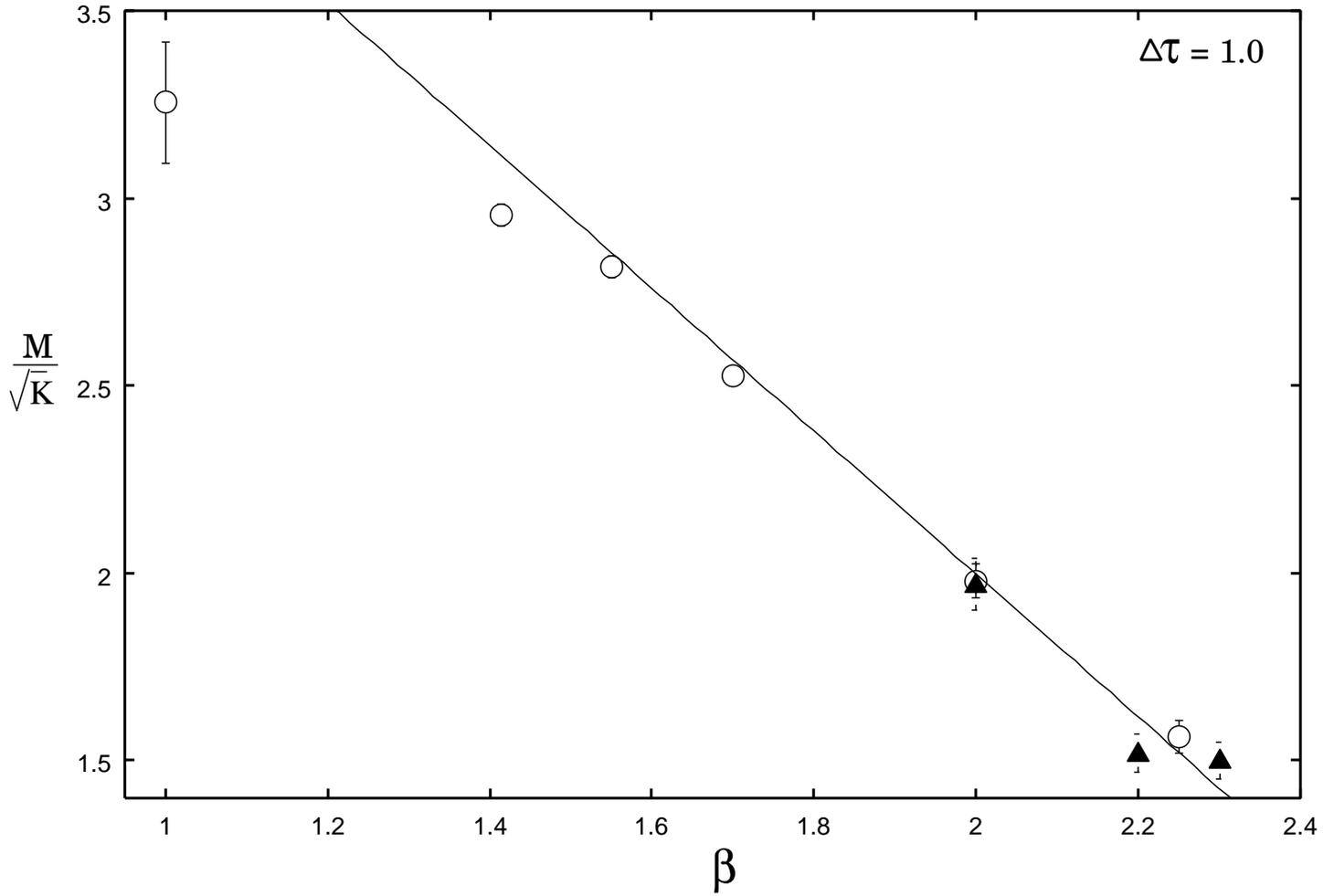}}
\caption{The dimensionless ratio $M/\sqrt{K}$ as a function
$\beta$. Our estimates are shown by circles and solid triangles
 show the earlier results of Teper
\protect\cite{tep99}. The solid curve
represents the  predicted weak-coupling behaviour.}
\label{fig3}
\end{figure}

To check the consistency of our method,
we plot the dimensionless ratio of the
the antisymmetric mass gap
over the square root of the string tension against $\beta$ together with
the results of Teper \cite{tep99} in Figure \ref{fig3}. The agreement is
excellent. The solid line gives the ratio of the fits in Figures
\ref{fig2} and Fig. (8) in \cite{loanb}, and shows how this ratio vanishes
exponentially in the weak-coupling limit, whereas in four-dimensional
confining theories it goes to a constant.
    
\begin{figure}
\centerline{\psfig{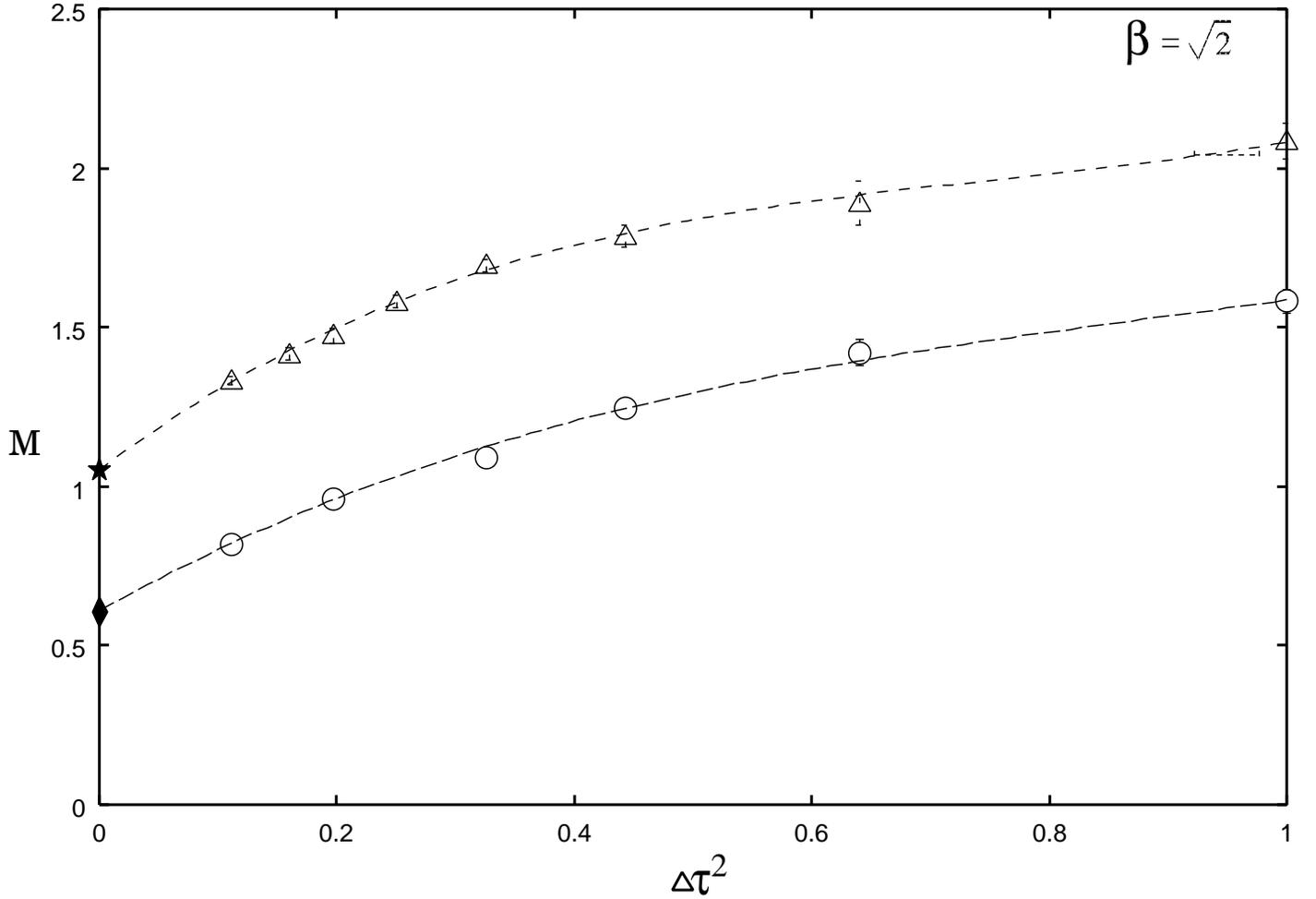}}
\caption{Estimates of the masses of $0^{++}$ and $0^{--}$ glueballs
against $\Delta \tau^{2}$. Results at $\beta =
\sqrt{2}$  for the $0^{++}$ and $0^{--}$ are labeled by
circles and triangle respectively.
The solid and dashed curves are the cubic fits to the data  extrapolated 
to
the Hamiltonian limit. The
series estimates of Hamer et al \protect\cite{ham92} in the limit
$\Dtau \rightarrow 0$, for symmetric and antisymmetric channels are
shown as a star and diamond respectively.}
\label{fig4}
\end{figure}

\begin{figure}
\centerline{\psfig{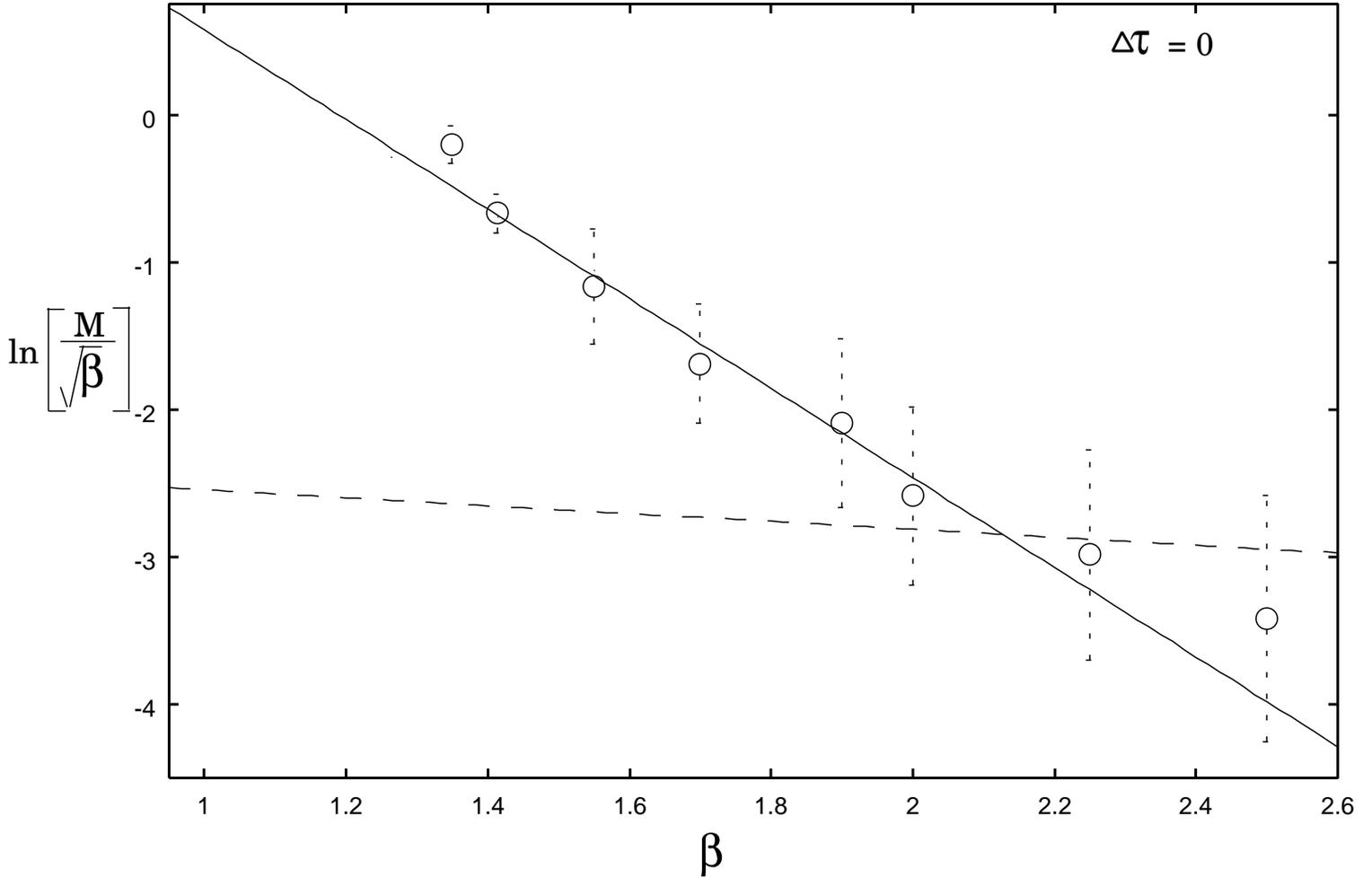}}
\caption{Hamiltonian estimates  of the  antisymmetric mass gap
plotted
as a function of $\beta$. The
solid curve is the fit to the data for $1.4<\beta <2.25$. The dashed line
represents the finite size effects \protect\cite{wei99}.}
\label{fig5}
\end{figure}

\begin{figure}
\centerline{\psfig{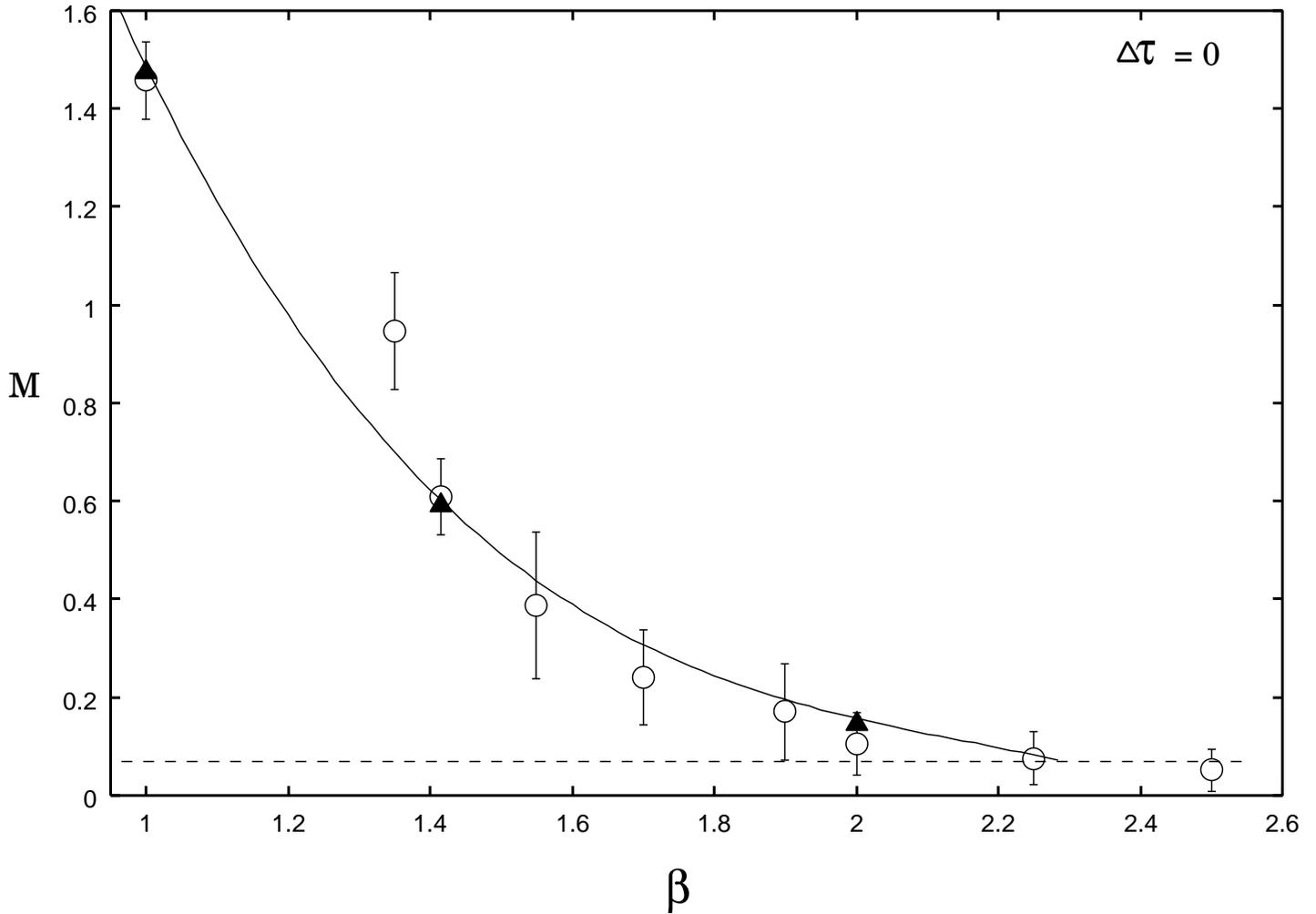}}
\caption{Hamiltonian estimates  of the  antisymmetric mass gap
plotted
as a function of $\beta$. The
solid curve represents the previous
results from series expansion \protect\cite{ham92} and 
the dashed line
represents the finite size effects \protect\cite{wei99}. The previous
 quantum Monte Carlo
calculations \protect\cite{ham94} are shown as solid triangles.}
\label{fig6}
\end{figure}

Figure \ref{fig4} shows the behaviour of the glueball masses as function
of ${\Delta\tau}^{2}$ for $\beta =\sqrt{2}$.
The extrapolation to the Hamiltonian
limit is performed by a simple cubic fit in powers of $\Dtau^{2}$. In the
limit $\Dtau \rightarrow 0$, we
reproduce the earlier  series estimates of Hamer et al\cite{ham92} for the
$0^{++}$ and $0^{--}$ states.  
The estimates of the antisymmetric mass gap, for various $\beta$, 
extracted from
the extrapolation to
the $\Dtau \rightarrow 0$ limit are  graphed as a function of $\beta$ in 
Figure \ref{fig5}. The solid line is a fit to the data for 
$1.4 \leq \beta \leq 2.25$ of
the form (\ref{eqn5}), with $v(0) = 0.3214$
and $c_1 = 5.50\pm 0.24$, which is similar to the coefficient found in
the                                                                    
Euclidean case. The fitting parameters, the slope, $3.10\pm 0.16$ and the intercept,
$3.61\pm 0.25$,  of 
the scaling curve are in excellent agreement with the results
obtained from the other Hamiltonian studies and are listed in table 
\ref{tablestat}.
The dashed line represents the finite-size scaling behaviour, equation
(\ref{eqn8}), which we assume holds in the Hamiltonian limit also, for
want of better information. It can be seen that the finite-size
corrections are predicted to dominate for $\beta \geq 2.2$, but the date
are not accurate enough at weak couplings to establish whether this is
really the case. 
Figure \ref{fig6}  shows the our Monte Carlo estimates for antisymmetric 
mass gap as a function of $\beta$. Also are shown the 
 results from previous strong-coupling series 
extrapolations\cite{ham92}
and quantum Monte Carlo calculations\cite{ham94}. It can be seen that our
present results agree with previous estimates but are less accurate.

\begin{figure}
\centerline{\psfig{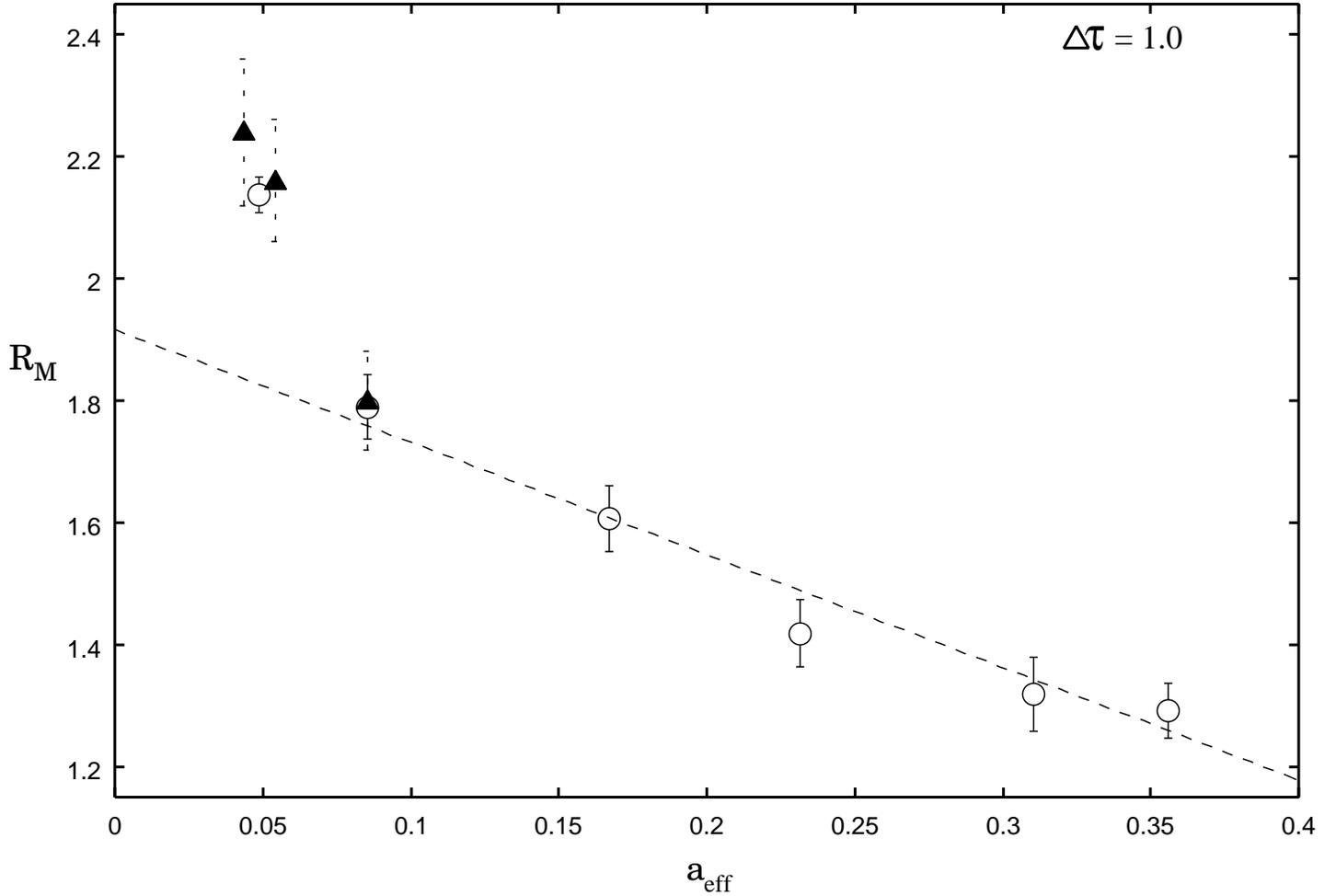}}
\caption{A graph showing estimates of the mass ratio $R_M$as a function of 
the effective spacing, $a_{eff}$, at $\Delta \tau =1.0$. Our present 
estimates are shown by the circles. The dashed line is a linear fit to 
the data over the  range $0.08\leq a_{eff} \leq 0.35$. The solid triangles 
show the previous estimates of Teper \protect\cite{tep99}.}
\label{fig7}
\end{figure}

Finally,
Figure \ref{fig7} displays the behaviour of the dimensionless mass
ratio,
\begin{equation}
R_M = M(0^{++})/M(0^{--})
\label{eqn10}
\end{equation}
for the Euclidean case $\Delta \tau = 1$. As in the (3+1)D confining
theories, we may expect that quantities of this sort will approach their
weak-coupling or continuum limits with corrections of $O(1/a_{eff})$,
where $a_{eff}$ is the effective lattice spacing in `physical' units
when the mass gap has been renormalized to a constant. Hence for our
present purposes we {\it define} $a_{eff}$ from equation (\ref{eqn5})
as
\begin{equation}
a_{eff} = \sqrt{8\pi^2\beta}\exp(-\pi^2\beta v(0))
\label{eqn11}
\end{equation}
with $v(0) = 0.2527$ for the Euclidean case. The mass ratio is plotted        
against $a_{eff}$ in Figure \ref{fig6}.
At weak coupling, we expect the theory to
approach a theory of free bosons\cite{gop82} so that the symmetric state 
will
be
composed of two $0^{--}$ bosons and the mass ratio should approach
two. Our results show that as
$a_{eff}$ goes to zero, the mass ratio rises to around the expected value 
of
$2.0$. A
linear fit to the data from $0.08 \leq a_{eff} \leq 0.35$ gives an
intercept     
However, we note that the last one of our estimates, together with two
from Teper\cite{tep99}, lie considerably {\it above} $R_M = 2$.
In the bluk systems, of course, the ratio cannot rise above 2, because
it is always possible to construct a $0^{++}$ state out of two $0^{--}$
mesons.    
been included in the fit.  

\section*{Acknowledgments} This work was supported by the Australian
Research Council. We are grateful for access to the
computing facilities of the Australian Centre for Advanced Computing and
Communications (ac3) and the Australian Partnership for Advanced
Computing (APAC).

\begin{table}
\begin{center}
\begin{tabular}{lll}
Source & $c_{0}$ & $c_{1}$ \\ \hline
Villain (Hamiltonian) & 3.17 & 2.18\\
Lana \protect\cite{lana88}    & 2.05 & 2.49 \\
Hamer and Irving \protect\cite{ham85}& 2.65 & 3.07\\
Hamer, Oitmaa and Weihong \protect\cite{ham92} & 2.71 & 3.13\\
Dabrighaus, Ristig and Clark \protect\cite{dab91}& 2.40 & 3.13\\
Fang, Liu and Guo \protect\cite{fan96}& 2.50 & 2.94\\
Morningstar \protect\cite{mor92} & 2.61 & 2.97\\
Heys and Stump \protect\cite{hey87} & 2.48 & 3.13\\
McIntosh and Hollenberg \protect\cite{john97} & 2.50 & 2.91 \\
Xiyan, Jinming and Shuohong \protect\cite{xiy96} & 2.50 & 2.95\\
Darooneh and Modarres \protect\cite{dar02}& $2.20\pm 0.03$ & $2.96\pm 0.05$\\
This work & $3.10\pm 0.16$ & $3.61\pm 0.25$  \\
\end{tabular}
\end{center}
\caption{Comparison of the coefficients $c_{0}$ ( the negative of
the slope of scaling curve) and $c_{1}$ (intercept of the scaling curve
on the $\mbox{ln}(M/\sqrt{\beta})$ axis) in the weak coupling limit
for antisymmetric mass gap.}
\label{tablestat}
\end{table}

%=======================================================================
%=FIGURES===============================================================
%=======================================================================
\center
\widetext
\input psfig
\psfull

%=======================================================================

%=======================================================================    

%=======================================================================
%=======================================================================
%=======================================================================
%=======================================================================

\end{document}